\documentclass[review]{elsarticle}
\usepackage{lineno,hyperref}
\usepackage{graphicx}
\usepackage{makecell}
\usepackage{color}
\usepackage{caption}
\usepackage{multirow}
\modulolinenumbers[5]
\journal{Journal of \LaTeX\ Templates}

\begin{document}
\captionsetup[figure]{labelfont={bf},labelformat={default},labelsep=period,name={Fig.}}

\begin{frontmatter}
\title{Link prediction via controlling the leading eigenvector}
\author[]{Yan-Li Lee}
\author[]{Qiang Dong}
\author[]{Tao Zhou\corref{cor1}}
\cortext[cor1]{Corresponding author.}
\ead{zhutou@ustc.edu}
\address{CompleX Lab, University of Electronic Science and Technology of China, Chengdu 611731, P.R. China}

\begin{abstract}
Link prediction is a fundamental challenge in network science. Among various methods, similarity-based algorithms are popular for their simplicity, interpretability, high efficiency and good performance. In this paper, we show that the most elementary local similarity index Common Neighbor (CN) can be linearly decomposed by eigenvectors of the adjacency matrix of the target network, with each eigenvector’s contribution being proportional to the square of the corresponding eigenvalue. As in many real networks, there is a huge gap between the largest eigenvalue and the second largest eigenvalue, the CN index is thus dominated by the leading eigenvector and much useful information contained in other eigenvectors may be overlooked. Accordingly, we propose a parameter-free algorithm that ensures the contributions of the leading eigenvector and the secondary eigenvector the same. Extensive experiments on real networks demonstrate that the prediction performance of the proposed algorithm is remarkably better than well-performed local similarity indices in the literature. A further proposed algorithm that can adjust the contribution of leading eigenvector shows the superiority over state-of-the-art algorithms with tunable parameters for its competitive accuracy and lower computational complexity.
\end{abstract}

\begin{keyword}
\texttt Complex Networks\sep Link Prediction\sep Similarity Index
\end{keyword}
\end{frontmatter}

\section{Introduction}

The booming of network science has brought about a new vision to explore and tackle problems existed in biology \cite{gosak2018network,barabasi2011network,barabasi2004network}, economics \cite{jackson2014networks,schweitzer2009economic,gao2019computational,lu2012recommender}, social science \cite{kossinets2006empirical,hu2019segregation}, data science \cite{zanin2016combining,stanisz2019linguistic,wu2013data} and so on. As an increasing number of datasets are structured by networks (e.g., protein-protein interaction networks, miRNA-disease association networks, drug-target networks, user-product networks, company investment networks, collaboration networks, citation networks, etc.), link prediction \cite{lu2011link,zhou2021progresses} has found wide applications in many scenarios. For example, in laboratorial experiments, instead of blindly checking all possible interactions among proteins, predicting potential interactions based on existed interactions can sharply reduce experimental costs if the prediction is accurate enough \cite{Ding2013similarity}. In social websites, academic websites and e-commerce platforms, a good recommendation of friends, citations, collaborators and products can enhance the loyalties and experience of users and the conversion rate in trading platforms \cite{lu2012recommender}. In sentiment analysis, an effective link prediction algorithm can well capture the relevance of different sentences \cite{martinvcic2017link}. In addition, link prediction is also helpful in probing network evolution mechanisms \cite{wang2012evaluating,zhang2015measuring} and estimating the extent of the regularity of the network formation \cite{lu2015toward,xian2020netsre}. In a word, link prediction is a significant and challenging problem in network science.

Various methods are proposed, including similarity-based algorithms \cite{liben2007link,zhou2009predicting,cannistraci2013link}, probabilistic models \cite{neville2007relational,yu2006stochastic}, maximum likelihood methods \cite{clauset2008hierarchical,guimera2009missing,pan2016predicting}, network embedding \cite{grover2016node2vec,tang2015line,wang2016structural} and other representatives \cite{pech2017link,benson2018simplicial,ghasemian2020stacking,jalili2017link}. Among them, similarity-based algorithms are popular for its simplicity, interpretability, high efficiency and satisfactory performance. In this paper, we show that the most elementary index, namely the common neighbor (CN) index \cite{liben2007link}, is dominated by the leading eigenvector of the adjacency matrix $\mathbf{A}$ of the target network, which is caused by the huge gap between the largest eigenvalue and the second largest eigenvalue in many real networks \cite{Farkas2001Spectra}. We propose a parameter-free algorithm that adopts a recently proposed enhancement framework for local similarity indices \cite{Lee2021} and ensures the contributions of the leading eigenvector and the secondary eigenvector the same. This algorithm is abbreviated as CLE because its underlying idea is controlling the leading eigenvector. Extensive experiments on real networks demonstrate that the prediction performance of CLE is remarkably better than well-performed local similarity indices and their enhanced versions. We further propose a parameter-dependent algorithm with adjustable contribution of the leading eigenvector, which outperforms elaborately designed algorithms that also have tunable parameters.

\section{Methods}

\begin{figure}[t]
\setlength{\abovecaptionskip}{0pt}
\centering
	\includegraphics[width=1.05\textwidth]{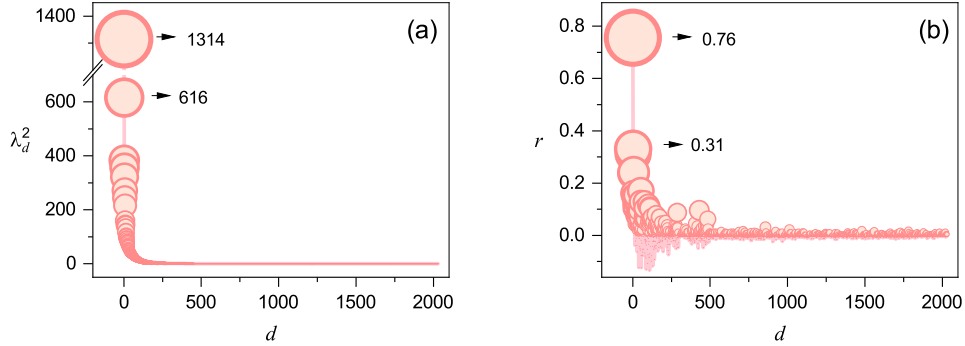}
    \caption{The dominant role of the leading eigenvector on DNC. (a) The relation between $\lambda_d^2$ and $d$, and (b) the Pearson correlation coefficient $r$ between $\mathbf{S}^{CN}$ and $\mathbf{S}^{(d)}$. The size of a circle is proportional to its corresponding value in $y$-axis. There exists a huge gap between $\lambda_1^2$ and $\lambda_2^2$, and thus the correlation coefficient $r$ between $\mathbf{S}^{CN}$ and $\mathbf{S}^{(1)}$ is larger than the correlation between $\mathbf{S}^{CN}$ and $\mathbf{S}^{(d)}$ for $d>1$.}
\label{fig.1}
\end{figure}

The CN index directly counts the number of common neighbors between two nodes, as
\begin{equation}
\label{eq.1}
S_{xy}^{CN}=|\Gamma_x\cap\Gamma_y|,
\end{equation}
where $x$ and $y$ are two arbitrary nodes and $\Gamma_x$ is the set of neighbors of $x$. The CN matrix $\mathbf{S}^{CN}$ can be decomposed as
\begin{equation}
\label{eq.2}
\mathbf{S}^{CN}=\mathbf{A}^2=\sum_{d=1}^{N}\lambda_d^2\mathbf{v}_d\mathbf{v}_d^T=\sum_{d=1}^{N}\lambda_d^2\mathbf{S}^{(d)},
\end{equation}
where $N$ is the number of nodes, $\lambda_d$ and $\mathbf{v}_d$ are the $d$th eigenvalue and the corresponding orthogonal and normalized eigenvector of $\mathbf{A}$, and $\lambda_1^2 \geq \lambda_2^2 \geq \cdots \geq \lambda_N^2$. Notice that, node $x$ can be represented by the $N$ eigenvectors as
\begin{equation}
\label{eq.3}
\mathbf{w}_x = [\lambda_1v_{1x}, \lambda_2v_{2x}, \cdots, \lambda_Nv_{Nx}],
\end{equation}
where $v_{dx}$ is the $x$th element of the $d$th eigenvector. Then, $\mathbf{S}^{CN}_{xy}$ is the dot product of $\mathbf{w}_x$ and $\mathbf{w}_y$, as
\begin{equation}
\label{eq.4}
S^{CN}_{xy}=|\Gamma_x\cap\Gamma_y|=\mathbf{w}_x\mathbf{w}_y^T=\sum_{d=1}^{N}\lambda_d^2v_{dx}v_{dy}.
\end{equation}
Accordingly, the contribution of each eigenvector to CN is proportional to the square of its eigenvalue. Therefore, the gap between $\lambda_1^2$ and $\lambda_2^2$ for many real networks \cite{Farkas2001Spectra} leads to the dominant contribution of the leading eigenvector. Take the email communication network DNC \cite{kunegis2013konect} as an example (the detailed description of this network is shown later), $\lambda_1^2$ of DNC is 2.13 times larger than $\lambda_2^2$. As a result, the Pearson correlation coefficient $r$ between $\mathbf{S}^{CN}$ and $\mathbf{S}^{(1)}$ is much higher than correlation coefficients between $\mathbf{S}^{CN}$ and others (see Fig.~\ref{fig.1}).

\begin{table*}[ht]
\footnotesize
\caption{Structural statistics of 18 real networks. $N$ and $M$ are the number of nodes and links. $\rho$, $\left\langle k \right\rangle$, $\left\langle c \right\rangle$, $\left\langle l \right\rangle$ and $\sigma$ are the network density, average degree, average clustering coefficient \cite{watts1998collective}, average shortest path length, and assortativity coefficient \cite{newman2002assortative}, respectively. $\delta=\lambda_2^2/\lambda_1^2$ measures the eigenvalue gap.}
\label{tab.1}
\begin{center}
\begin{tabular}{|p{1.75cm}<{\centering}||cccccccc|}
\hline
\rule{0pt}{9pt}
Network & $N$ & $M$ & $\rho$ & $\left\langle k \right\rangle$ & $\left\langle c \right\rangle$ &  $\left\langle l \right\rangle$ & $\sigma$ & $\delta$\\
\hline\hline
FWF&128&2106&0.259&32.91 &0.33&1.77&-0.10&0.267\\
FWE&69&880&0.375&25.51 &0.55&1.64&-0.27&0.195\\
RAD&167&3250&0.234&38.92 &0.59&1.97&-0.30&0.086\\
DNC&2029&4384&0.002&4.32 &0.22&3.37&-0.31&0.468\\
HFR&1858&12534&0.007&13.49 &0.14&3.45&-0.09&0.283\\
HG&274&2124&0.057&15.50 &0.63&2.42&-0.47&0.075\\
WR&6875&64712&0.003&18.83 &0.30&3.49&-0.24&0.544\\
PH&241&923&0.068&7.66 &0.22&2.59&-0.08&0.992\\
MR&1682&94834&0.067&112.76 &0.36&2.16&-0.19&0.118\\
BG&1224&16715&0.022&27.31 &0.32&2.74&-0.22&0.655\\
GFA&297&2148&0.046&14.04 &0.29&2.45&-0.16&0.342\\
BM13&3391&4388&0.001&2.59 &0.07&6.61&-0.02&0.895\\
FG&2239&6432&0.003&5.75 &0.04&3.84&-0.33&0.910\\
FTB&35&118&0.193&6.74 &0.27&2.13&-0.26&0.352\\
WTN&80&875&0.277&21.88 &0.75&1.72&-0.39&0.184\\
UST&332&2126&0.039&12.81 &0.62&2.74&-0.21&0.176\\
ATC&1266&2408&0.003&3.93 &0.07&5.93&-0.02&0.723\\
ER&1174&1417&0.002&2.41 &0.02&18.40&0.09&0.955\\
\hline
\end{tabular}
\end{center}
\end{table*}

The dominant role of the leading eigenvector may oversuppress useful information contained in other eigenvectors, and thus to depress the contribution of the leading eigenvector may better characterize the node similarity. The most straightforward way to tackle this problem is to make the contribution of $\mathbf{S}^{(1)}$ the same as $\mathbf{S}^{(2)}$. Accordingly, a parameter-free similarity matrix can be obtained as
\begin{equation}
\label{eq.5}
\mathbf{\tilde{S}}=\lambda_2^2\mathbf{S}^{(1)}+\sum_{d=2}^{N}\lambda_d^2\mathbf{S}^{(d)}.
\end{equation}
It is natural to extend $\mathbf{\tilde{S}}$ to a parameter-dependent similarity matrix as
\begin{equation}
\label{eq.6}
\mathbf{\tilde{S}^*}=\alpha\lambda_1^2\mathbf{S}^{(1)}+\sum_{d=2}^{N}\lambda_d^2\mathbf{S}^{(d)},
\end{equation}
where $\alpha$ is a tunable parameter controlling the contribution of $\mathbf{v}_1$.

Very recently, an enhancement framework, named self-included collaborative filtering (SCF), is proposed for local similarity indices \cite{Lee2021}. Given a similarity matrix $\mathbf{S}$, the SCF-enhanced matrix reads
\begin{equation}
\label{eq.7}
\ddot{\mathbf{S}}=(\mathbf{A}+\mathbf{I})\mathbf{S}+[(\mathbf{A}+\mathbf{I})\mathbf{S}]^T,
\end{equation}
where $\mathbf{I}$ is the identity matrix. This framework largely improves the prediction performance and the robustness of local similarity indices. By applying the enhancement framework on $\mathbf{\tilde{S}}$ and $\mathbf{\tilde{S}^*}$, we can obtain a parameter-free algorithm (CLE) as
\begin{equation}
\label{eq.8}
\mathbf{S}^{CLE}=(\mathbf{A}+\mathbf{I})\mathbf{\tilde{S}}+[(\mathbf{A}+\mathbf{I})\mathbf{\tilde{S}}]^T
\end{equation}
and a parameter-dependent algorithm (CLE*) as
\begin{equation}
\label{eq.9}
\mathbf{S}^{CLE^*}=(\mathbf{A}+\mathbf{I})\mathbf{\tilde{S}^*}+[(\mathbf{A}+\mathbf{I})\mathbf{\tilde{S}^*}]^T.
\end{equation}
Clearly, CLE* reduces to CLE when $\alpha=\lambda_2^2/\lambda_1^2$, and to $\ddot{\mathbf{S}}^{CN}$ when $\alpha=1$ ($\ddot{\mathbf{S}}^{CN}$ is the SCF-enhanced CN index).

\section{Results}
To test the algorithmic performance, 18 networks from disparate fields are considered, including (1) FWF \cite{rossi2015network}--the predator-prey network of animals in the ecosystem of Coastal bay in Florida Bay in the dry season; (2) FWE \cite{rossi2015network}--the predator-prey network of animals in Everglades Graminoids in the west season; (3) RAD \cite{kunegis2013konect}--the email communication network between employees of a mid-sized manufacturing company; (4) DNC \cite{kunegis2013konect}--the email communication network in the 2016 Democratic Committee email leak; (5) HFR \cite{kunegis2013konect}--the friendship network among users of the website \emph{hamsterster.com}; (6) HG \cite{kunegis2013konect}--the human contact network measured by carried wireless devices; (7) WR \cite{hu2019segregation}--the religious network where each node represents a \emph{Weibo} user with explicit religious belief and each link denotes a follower-followee relationship; (8) PH \cite{kunegis2013konect}--the innovation spreading network in which nodes denote physicians and links denote friendships or discussions between physicians; (9) MR \cite{moviedata}--the rating network where links denote ratings of users to movies; (10) BG \cite{kunegis2013konect}--the hyperlink network among blogs in the context of the 2004 US election; (11) GFA \cite{watts1998collective}--the gene functional association network of C.elegans; (12) BM13 \cite{kovacs2019network}--the protein-protein interaction network in Lit-BM-13; (13) FG \cite{kunegis2013konect}--the protein-protein interaction network in Homo sapiens; (14) FTB \cite{footballdata}--the  trading network of players of 35 national soccer teams; (15) WTN \cite{de2011exploratory}--the world trading network of miscellaneous manufactures of metal among 80 countries in 1994; (16) UST \cite{kunegis2013konect}--the air transportation network of US; (17) ATC \cite{kunegis2013konect}--the network of airports or service centers, where each link denotes a preferred route between two nodes; (18) ER \cite{kunegis2013konect}--the international E-road network between cities in Europe where each link denotes a road between two cities. In the above networks, multiple links and self-connections are not allowed, and directions and weights of links are ignored. The elementary statistics are shown in Table~\ref{tab.1}.

Consider a network $G(V, E)$ with $V$ the set of nodes and $E$ the set of links. To evaluate the algorithmic performance, $E$ is randomly divided into two parts: the training set $E^T$ contains the known topology, and the testing set $E^P$ is treated as missing links that cannot be used in the prediction. Obviously, $E^T\cup E^P = E$ and $E^T\cap E^P = \emptyset$. Many link prediction algorithms will assign a score to each link and the top-$L$ links (i.e., the $L$ links with highest scores and not belonging to $E^T$) are the predicted links. Two metrics, AUC and AUPR, are used to evaluate the algorithmic performance \cite{hanley1982meaning, provost2001robust,lichtenwalter2010new,yang2015evaluating}. AUC denotes the probability that a randomly chosen missing link is assigned a higher score than a randomly chosen nonexistent link. To calculate AUC, we randomly select $n$ pairs of links respectively from $E^P$ and $U-E$, where $U$ is the universal set containing all the $N(N-1)/2$ potential links. If there are $n_1$ times the missing link has higher score and $n_2$ times the missing link has the same score compared to the nonexistent link, the corresponding AUC is $(n_1+0.5n_2)/n$. If all scores of the candidate links are randomly generated from an independent and identical distribution, the AUC value should be 0.5, and the value exceeds 0.5 indicates the extent that the considered algorithm outperforms the pure chance. AUPR is the area under the precision-recall curve \cite{yang2015evaluating}: the larger the value, the better the performance. Precision is defined as the ratio of relevant links selected to the number of links selected. If $L_r$ links among the top-$L$ selected links are correctly predicted, precision is $L_r/L$. Recall is defined as the ratio of relevant links selected to the total number of relevant links, say $L_r/|E^P|$. Compared to the direct usage of precision and recall, AUPR does not depend on the choice of $L$ and shows a more comprehensive evaluation by considering a range of $L$.

Eight parameter-free local similarity indices are used to compare with CLE, including the CN index \cite{liben2007link}, the Adamic-Adar (AA) index \cite{adamic2003friends}, the resource allocation (RA) index  \cite{zhou2009predicting}, the CRA index \cite{cannistraci2013link} and their SCF-enhanced versions. AA differentiates the role of common neighbors by depressing the contributions of large-degree nodes, as
\begin{equation}
\label{eq.10}
S_{xy}^{AA}=\sum_{z\in\Gamma_x\cap\Gamma_y}\frac{1}{\log k_z},
\end{equation}
where $k_z$ is the degree of node $z$. RA quantifies node similarity by a resource allocation process \cite{ou2007power}, where each neighbor of $x$ allocates one unit resource equally to its neighbors, and the similarity of $x$ and $y$ is the resource that $y$ receives from $x$, namely
\begin{equation}
\label{eq.11}
S_{xy}^{RA}=\sum_{z\in\Gamma_x\cap\Gamma_y}\frac{1}{k_z}.
\end{equation}
CRA prefers node pairs with densely connected common neighbors \cite{cannistraci2013link}, as
\begin{equation}
\label{eq.12}
S_{xy}^{CRA}=\sum_{z\in\Gamma_x\cap\Gamma_y}\frac{\gamma_z}{k_z},
\end{equation}
where $\gamma_z=|\Gamma_z\cap\Gamma_x\cap\Gamma_y|$. The enhanced versions of CN, AA, RA and CRA (i.e., $\ddot{\mathbf{S}}^{CN}$, $\ddot{\mathbf{S}}^{AA}$, $\ddot{\mathbf{S}}^{RA}$ and $\ddot{\mathbf{S}}^{CRA}$) can be obtained by Eq.~(\ref{eq.7}).

AUC and AUPR of 9 indices are reported in Table~\ref{tab.2} and Table~\ref{tab.3}. Of course, no algorithm can be superior than all other algorithms for all networks \cite{ghasemian2020stacking}, thereby we further design three derived evaluation metrics to evaluate the overall performance of algorithms on all networks. They are the winning rate $R$, the average AUC value and the average AUPR value. $R$ values of 9 indices are calculated based on 18 comparisons on 18 real networks. In each comparison, the best-performed index will get score 1, and if $m>1$ indices are equally best, they all get score 1/$m$. The $R$ value of each index is its total score divided by the number of comparisons. Overall speaking, CLE is remarkably better than all other indices according to the winning rate values $R$, the average AUC values and the average AUPR values, and CLE achieves significant improvement compared with $\ddot{\mathbf{S}}^{CN}$ by controlling the contribution of $\mathbf{v}_1$.

\begin{table*}[hbtp]
\footnotesize
\caption{AUC of CLE and 8 benchmark indices on 18 real networks. Each result is averaged over 100 independent runs with $E^P$ containing 10\% links. The best-performed result for each network, the highest winning rate $R$ and the highest average AUC value are emphasized in bold. CLE performs overall best among all indices considered.}
\label{tab.2}
\begin{center}
\begin{tabular}{|p{1.2cm}<{\centering}||p{0.7cm}<{\centering}p{0.7cm}<{\centering}p{0.7cm}<{\centering}p{0.7cm}<{\centering}p{0.7cm}<{\centering}p{0.7cm}<{\centering}p{0.7cm}<{\centering}p{0.7cm}<{\centering}p{0.7cm}<{\centering}|}
\hline
\rule{0pt}{9pt}
Network  & CLE & $\ddot{\mathbf{S}}^{CN}$&$\ddot{\mathbf{S}}^{AA}$ &$\ddot{\mathbf{S}}^{RA}$&$\ddot{\mathbf{S}}^{CRA}$&$\mathbf{S}^{CN}$&$\mathbf{S}^{AA}$ &$\mathbf{S}^{RA}$&$\mathbf{S}^{CRA}$\\
\hline\hline
FWF&\textbf{0.9103} &0.8173 &0.8172 &0.8299 &0.8480 &0.6047 &0.6009 &0.6078 &0.6367 \\
FWE&\textbf{0.9219} &0.8457 &0.8491 &0.8609 &0.8890 &0.6758 &0.6876 &0.6923 &0.7060 \\
RAD&\textbf{0.9412} &0.9093 &0.9109 &0.9150 &0.8866 &0.9130 &0.9128 &0.9173 &0.9170 \\
DNC&\textbf{0.8481} &0.7899 &0.7904 &0.7922 &0.7696 &0.8004 &0.7998 &0.8054 &0.7486 \\
HFR&0.9520 &0.9353 &0.9445 &\textbf{0.9546} &0.9096 &0.8046 &0.8073 &0.8074 &0.6520 \\
HG&\textbf{0.9503} &0.9322 &0.9340 &0.9350 &0.9038 &0.9322 &0.9318 &0.9333 &0.9250 \\
WR&\textbf{0.9737} &0.9638 &0.9652 &0.9690 &0.9543 &0.9250 &0.9283 &0.9290 &0.8714 \\
PH&0.9008 &0.9082 &0.9069 &\textbf{0.9098} &0.7695 &0.8447 &0.8409 &0.8415 &0.6584 \\
MR&\textbf{0.9428} &0.9203 &0.9205 &0.9251 &0.9216 &0.9042 &0.9038 &0.9034 &0.9061 \\
BG&0.9398 &0.9308 &0.9340 &\textbf{0.9404} &0.9234 &0.9189 &0.9210 &0.9225 &0.8960 \\
GFA&\textbf{0.8977} &0.8505 &0.8677 &0.8832 &0.8250 &0.8482 &0.8665 &0.8707 &0.7644 \\
BM13&\textbf{0.7210} &0.6839 &0.6836 &0.6839 &0.5384 &0.5906 &0.5923 &0.5890 &0.5155 \\
FG&\textbf{0.9062} &0.8525 &0.8576 &0.8566 &0.6972 &0.5502 &0.5538 &0.5556 &0.5101 \\
FTB&\textbf{0.8072} &0.7492 &0.7764 &0.7740 &0.7500 &0.6498 &0.6535 &0.6311 &0.6687 \\
WTN&\textbf{0.9260} &0.8791 &0.8845 &0.8959 &0.8367 &0.8575 &0.8762 &0.8978 &0.8906 \\
UST&0.9438 &0.9020 &0.9113 &0.9304 &0.8794 &0.9322 &0.9471 &\textbf{0.9519} &0.9191 \\
ATC&\textbf{0.7559} &0.7154 &0.7155 &0.7184 &0.5300 &0.6112 &0.6132 &0.6091 &0.5095 \\
ER&\textbf{0.5544} &0.5514 &0.5521 &0.5540 &0.5000 &0.5260 &0.5232 &0.5244 &0.5000 \\
\hline\hline
$R$ &\textbf{77.78\%} &0.00\%&0.00\%&16.67\%&0.00\%&0.00\%&0.00\%&5.56\%&0.00\%\\
$\langle AUC \rangle$&\textbf{0.8774}&0.8409&0.8456&0.8516&0.7962&0.7716&0.7756&0.7772&0.7331\\
\hline
\end{tabular}
\end{center}
\end{table*}

\begin{table*}[hbtp]
\footnotesize
\caption{AUPR of CLE and 8 benchmark indices on 18 real networks. Each result is averaged over 100 independent runs with $E^P$ containing 10\% links. The best-performed result for each network, the highest winning rate $R$ and the highest average AUPR value are emphasized in bold. CLE performs overall best among all indices considered.}
\label{tab.3}
\begin{center}
\begin{tabular}{|p{1.2cm}<{\centering}||p{0.7cm}<{\centering}p{0.7cm}<{\centering}p{0.7cm}<{\centering}p{0.7cm}<{\centering}p{0.7cm}<{\centering}p{0.7cm}<{\centering}p{0.7cm}<{\centering}p{0.7cm}<{\centering}p{0.7cm}<{\centering}|}
\hline
\rule{0pt}{9pt}
Network  & CLE & $\ddot{\mathbf{S}}^{CN}$&$\ddot{\mathbf{S}}^{AA}$ &$\ddot{\mathbf{S}}^{RA}$&$\ddot{\mathbf{S}}^{CRA}$&$\mathbf{S}^{CN}$&$\mathbf{S}^{AA}$ &$\mathbf{S}^{RA}$&$\mathbf{S}^{CRA}$\\
\hline\hline
FWF&\textbf{0.4344} &0.2196 &0.2202 &0.2199 &0.3025 &0.0498 &0.0504 &0.0505 &0.0588 \\
FWE&\textbf{0.5163} &0.3131 &0.3167 &0.3223 &0.4650 &0.1223 &0.1281 &0.1354 &0.1282 \\
RAD&\textbf{0.5445} &0.3901 &0.3956 &0.4145 &0.2647 &0.4081 &0.4176 &0.4307 &0.4328 \\
DNC&\textbf{0.1428} &0.1005 &0.1068 &0.1120 &0.0728 &0.1217 &0.1190 &0.0913 &0.1247 \\
HFR&\textbf{0.2648} &0.0590 &0.0672 &0.0996 &0.1071 &0.0121 &0.0126 &0.0128 &0.0111 \\
HG&0.5839 &0.6313 &0.6351 &\textbf{0.6437} &0.5173 &0.5834 &0.5819 &0.5809 &0.5733 \\
WR&\textbf{0.1062} &0.0725 &0.0789 &0.0895 &0.0722 &0.0524 &0.0547 &0.0507 &0.0741 \\
PH&0.0521 &0.0514 &0.0540 &0.0562 &0.0391 &0.0564 &\textbf{0.0619} &0.0595 &0.0460 \\
MR&\textbf{0.1972} &0.1210 &0.1212 &0.1214 &0.1347 &0.0864 &0.0853 &0.0773 &0.0947 \\
BG&\textbf{0.1209} &0.1137 &0.1146 &0.1134 &0.0927 &0.0959 &0.0921 &0.0757 &0.1044 \\
GFA&\textbf{0.0910} &0.0501 &0.0526 &0.0572 &0.0472 &0.0479 &0.0530 &0.0530 &0.0575 \\
BM13&0.0109 &0.0094 &0.0150 &\textbf{0.0183} &0.0054 &0.0025 &0.0037 &0.0036 &0.0046 \\
FG&0.0412 &0.0406 &0.0423 &\textbf{0.0432} &0.0212 &0.0008 &0.0009 &0.0010 &0.0018 \\
FTB&\textbf{0.1555} &0.1323 &0.1372 &0.1395 &0.1209 &0.0639 &0.0592 &0.0552 &0.1048 \\
WTN&\textbf{0.4741} &0.3681 &0.3745 &0.3855 &0.2810 &0.3458 &0.3725 &0.3906 &0.3940 \\
UST&0.4119 &0.3320 &0.3411 &0.3652 &0.2535 &0.3455 &0.3711 &\textbf{0.4221} &0.3613 \\
ATC&0.0073 &0.0063 &0.0076 &\textbf{0.0083} &0.0017 &0.0050 &0.0034 &0.0029 &0.0013 \\
ER&0.0006 &\textbf{0.0008} &0.0006 &0.0006 &0.0002 &0.0006 &0.0004 &0.0004 &0.0003 \\
\hline\hline
$R$&\textbf{61.10}\%&5.60\%&0.00\%&22.20\%&0.00\%&0.00\%&5.60\%&5.60\%&0.00\%\\
$\langle AUC\rangle$&\textbf{0.2309} &0.1673 &0.1712 &0.1784 &0.1555 &0.1334 &0.1371 &0.1385 &0.1430 \\
\hline
\end{tabular}
\end{center}
\end{table*}

Take a close look at the results, one can find that the performance of CLE is either the best (in 14 of the 18 cases) or very close to the best subject to AUC. Subject to AUPR, in 7 of 18 cases, CLE is not the best. Among the 7 cases, for HG, PH, BM13, ATC and ER, AUPR of the best-performed index is $>$10\% better than that of CLE. Since the eigenvalue gaps in PH, BM13, ATC and ER are very small (corresponding to large $\delta$, see Table~\ref{tab.1}), the effect from controlling the leading eigenvector may be insignificant resulting in unsatisfied performance of CLE. HG is a special case, which is sparse yet highly clustered ($\langle c \rangle$=0.6327). Its first-order null model (with same degree sequence obtained by link-crossing operations, see the method proposed in Ref. \cite{maslov2002specificity}) is of almost the same clustering coefficient ($\langle c \rangle$=0.6320), and thus we think the degree information largely determines the linking pattern. We have tested the preferential attachment (PA) index \cite{barabasi1999emergence}, which is
\begin{equation}
\label{eq.13}
S_{xy}^{PA}=k_xk_y.
\end{equation}
 It performs very badly in most networks but achieves even higher AUPR (0.6333, close to the best result) than CLE for HG. Due to the leading eigenvector is strongly correlated with degree vector, to weaken the contribution of leading eigenvector may not work well for HG.

 To be fair, we compare CLE* with 3 parameter-dependent benchmarks, including the Katz index \cite{katz1953new}, the local path (LP) index  \cite{lu2009similarity} and the linear optimization (LO) algorithm \cite{pech2019link}. The Katz index considers all paths with exponentially damped contributions along with their lengths, as
\begin{equation}
\label{eq.14}
S^{Katz}_{xy}=\beta(\mathbf{A})_{xy}+\beta^2(\mathbf{A}^2)_{xy}+\beta^3(\mathbf{A}^3)_{xy}+\cdots,
\end{equation}
where $\beta$ is a tunable parameter. To avoid the degeneracy of states in CN and make similarities more distinguishable, LP considers both contributions from 2-hop and 3-hop paths, as
\begin{equation}
\label{eq.15}
S_{xy}^{LP}=(\mathbf{A}^2)_{xy}+\epsilon(\mathbf{A}^3)_{xy},
\end{equation}
where $\epsilon$ is a tunable parameter. LO assumes that the existence likelihood of a link is a linear summation of contributions of all its neighbors. By solving a corresponding optimization function, the analytical expression of the predicted matrix (can be treated as a similarity matrix) is
\begin{equation}
\label{eq.16}
\mathbf{S}^{LO}=\mathbf{A}\mathbf{Z}^*=\alpha\mathbf{A}(\alpha\mathbf{A}^T\mathbf{A}+\mathbf{I})^{-1}\mathbf{A}^T\mathbf{A},
\end{equation}
where $\alpha$ is a free parameter.

\begin{table}[t]
\footnotesize
\caption{AUC and AUPR of CLE* and 3 parameter-dependent benchmarks on 18 real networks. Parameters are tuned to their optimal values corresponding to the highest AUC values and the highest AUPR values, respectively. Each result is averaged over 100 independent runs with $E^P$ containing 10\% links. The best-performed result for each network, the highest winning rate $R$, the highest average AUC value and the highest average AUPR value are emphasized in bold. CLE* performs overall best subject to AUC, and LO performs overall best subject to AUPR.}
\label{tab.4}
\begin{center}
\begin{tabular}{|m{1.3cm}<{\centering}||m{0.9cm}<{\centering}m{0.9cm}<{\centering}m{0.9cm}<{\centering}m{0.9cm}<{\centering}|m{0.9cm}<{\centering}m{0.9cm}<{\centering}m{0.9cm}<{\centering}m{0.9cm}<{\centering}|}
\hline
\multirow{2}{*}{Network}&CLE*&$\mathbf{S}^{LO}$&$\mathbf{S}^{LP}$&$\mathbf{S}^{Katz}$& CLE*&$\mathbf{S}^{LO}$&$\mathbf{S}^{LP}$&$\mathbf{S}^{Katz}$\\
\cline{2-5}\cline{6-9}
&\multicolumn{4}{c|}{AUC}&\multicolumn{4}{c|}{AUPR}\\
\hline\hline
FWF&0.9197&\textbf{0.9485}&0.8122&0.7134&0.4407&\textbf{0.5967}&0.2258&0.0975 \\
FWE&0.9235&\textbf{0.9359}&0.8447&0.7623&0.5494&\textbf{0.6252}&0.3216&0.1897 \\
RAD&\textbf{0.9445}&0.9411&0.9095&0.9109&\textbf{0.5500}&0.5443&0.4084&0.4115 \\
DNC&\textbf{0.8489}&0.8372&0.7876&0.7444&\textbf{0.1464}&0.1269&0.1194&0.1212 \\
HFR&\textbf{0.9534}&0.9515&0.9371&0.9113&0.2971&\textbf{0.5243}&0.0601&0.0228 \\
HG&\textbf{0.9567}&0.9467&0.9357&0.9280&0.6442&\textbf{0.6463}&0.6306&0.6235 \\
WR&\textbf{0.9738}&0.9714&0.9632&0.9517&0.1129&\textbf{0.1680}&0.0751&0.0588 \\
PH&0.9150&0.8893&0.9138&\textbf{0.9211}&0.0554&0.0481&\textbf{0.0622}&0.0583 \\
MR&0.9438&\textbf{0.9494}&0.9204&0.9084&0.2018&\textbf{0.2456}&0.1210&0.1075 \\
BG&0.9404&\textbf{0.9436}&0.9317&0.9273&0.1210&\textbf{0.1682}&0.1130&0.1018 \\
GFA&\textbf{0.9006}&0.8904&0.8700&0.8656&0.0920&\textbf{0.1046}&0.0544&0.0490 \\
BM13&\textbf{0.7266}&0.6378&0.6867&0.6410&0.0130&\textbf{0.0157}&0.0101&0.0045 \\
FG&\textbf{0.9153}&0.8947&0.8533&0.7746&0.0460&\textbf{0.0533}&0.0382&0.0089 \\
FTB&0.7860&\textbf{0.8092}&0.7522&0.7332&0.1616&\textbf{0.1656}&0.1473&0.1170 \\
WTN&\textbf{0.9277}&0.9273&0.8796&0.8699&\textbf{0.4750}&0.4744&0.3809&0.3750 \\
UST&\textbf{0.9498}&0.9385&0.9365&0.9234&\textbf{0.4258}&0.3890&0.3459&0.3399 \\
ATC&\textbf{0.7602}&0.6128&0.7147&0.7547&0.0076&\textbf{0.0086}&0.0066&0.0066 \\
ER&0.5664&0.5035&0.5537&\textbf{0.6239}&0.0008&0.0006&0.0008&\textbf{0.0014} \\
\hline\hline
$R$ &\textbf{61.11\%}&27.78\%&0.00\%&11.11\%&22.22\%&\textbf{66.67\%}&5.56\%&5.56\%\\
$\langle AUC \rangle$&\textbf{0.8820}&0.8617&0.8446&0.8258&0.2411&\textbf{0.2725}&0.1734&0.1497\\
\hline
\end{tabular}
\end{center}
\end{table}
\clearpage

The AUC values and the AUPR values of CLE* and three compared algorithms are reported in Table~\ref{tab.4}. CLE* outperforms all others subject to AUC, and is the runner-up subject to AUPR. Overall speaking, CLE* and LO (the latter is known to be one of the most accurate algorithms thus far) are competitive in accuracy, and they are remarkably better than LP and Katz indices.

\section{Analyses}

\begin{figure}[ht!]
\setlength{\abovecaptionskip}{0pt}
\centering
	\includegraphics[width=1\textwidth]{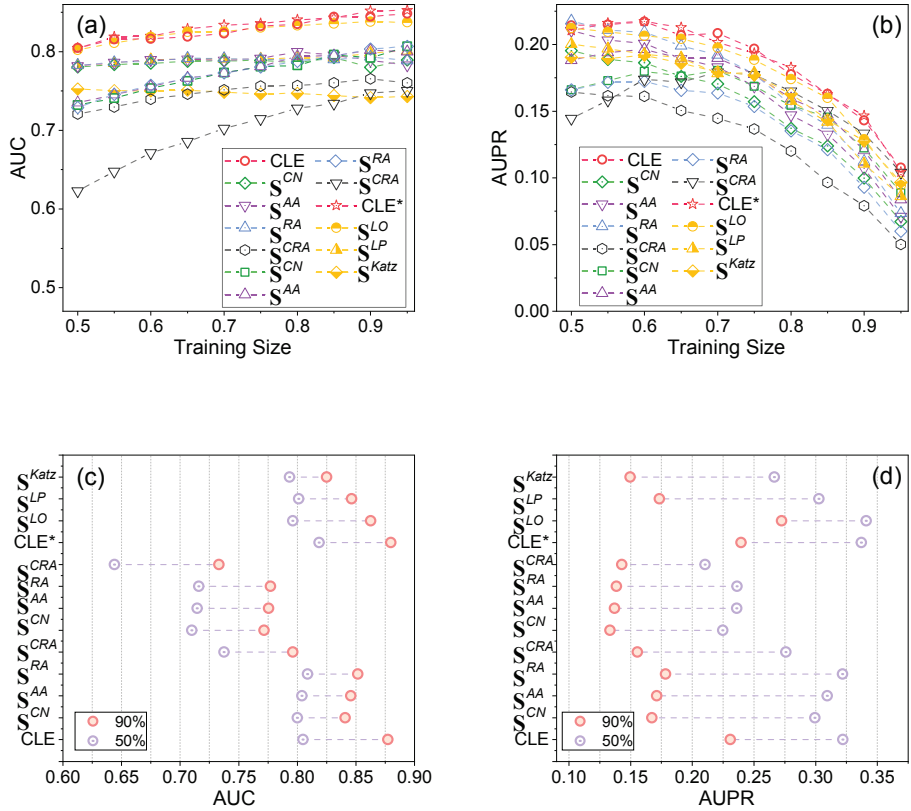}
    \caption{The robustness of CLE and CLE* against the data sparsity. (a) and (b) show AUC and AUPR of 13 algorithms for different training sizes on DNC. (c) and (d) show the average AUC values and the average AUPR values over 18 real networks for 13 algorithms, where the ratios of training set are 50\% and 90\%, respectively. The results are averaged over 100 independent runs, and parameters are tuned to their optimal values subject to the highest AUC values and the highest AUPR values respectively. CLE and CLE* are stable and overall best subject to AUC, and are competitive with LO subject to AUPR.}
\label{fig.2}
\end{figure}

We first test the robustness of the considered algorithms against the data sparsity by varying the size of the training set from 50\% to 95\%. Fig.~\ref{fig.2}(a) and Fig.~\ref{fig.2}(b) show the result for DNC. CLE and CLE* are relatively insensitive to the data sparsity subject to AUC, and they always perform best compared to other considered algorithms. Fig.~\ref{fig.2}(c) and Fig.~\ref{fig.2}(d) further show the average AUC values and the average AUPR values over 18 real networks with the ratios of training set being 50\% and 90\%, respectively. One can observe that CLE* performs best subject to AUC, while LO performs best subject to AUPR.

Next, we analyze the time complexity of the considered algorithms. To calculate CN, AA and RA, for each node $x$, we first search its neighbors, and then search the neighbors of neighbors, so that all 2-hop paths are taken into account. Therefore, the time complexity of CN, AA and RA is $O(N\left\langle k \right\rangle^2)$. The time complexity of CRA, $\ddot{\mathbf{S}}^{CN}$, $\ddot{\mathbf{S}}^{AA}$, $\ddot{\mathbf{S}}^{RA}$ and LP is $O(N\left\langle k \right\rangle^3)$ since one more search is required. Analogously, the time complexity of $\ddot{\mathbf{S}}^{CRA}$ is $O(N\left\langle k \right\rangle^4)$. The time complexity of the Katz index and LO is $O(N^3)$, dominated by the matrix inversion operator. Direct spectral decomposition in CLE and CLE* is time-consuming, instead we can rewrite $\mathbf{\tilde{S}}$ and $\mathbf{\tilde{S}^*}$ as
\begin{equation}
\label{eq.17}
\mathbf{\tilde{S}}= \mathbf{A}^2 + (\lambda_2^2-\lambda_1^2)\mathbf{S}^{(1)}
\end{equation}
and
\begin{equation}
\label{eq.18}
\mathbf{\tilde{S}^*}=\mathbf{A}^2+(\alpha-1)\lambda_1^2\mathbf{S}^{(1)}.
\end{equation}

\begin{figure}[t]
\setlength{\abovecaptionskip}{0pt}
\centering
	\includegraphics[width=0.65\textwidth]{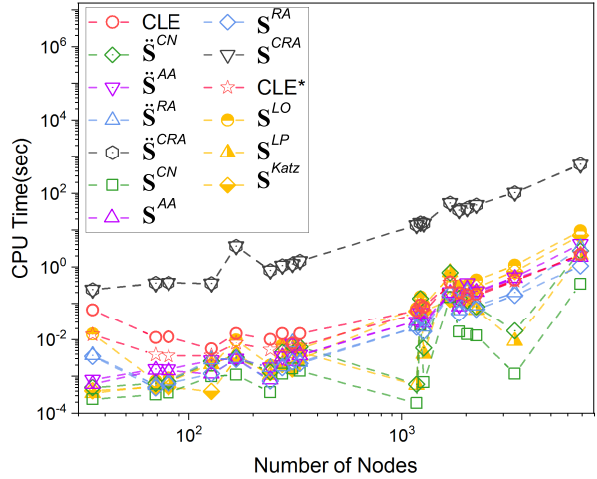}
    \caption{CPU times of different algorithms with varying network sizes. 18 data points for each algorithm correspond to 18 real networks. All computations are implemented on a desktop computer with a single Intel(R) Core (TM) processor (3.40 GHZ). Results for CLE and CLE* are marked in red. CPU times are averaged over 100 independent runs. CLE and CLE* take shorter CPU times than global algorithms.}
\label{fig.3}
\end{figure}
As a result, the time consumption of CLE and CLE* mainly comes from the calculation of the leading eigenvector, the largest eigenvalue and the second largest eigenvalue, which can be quickly obtained by the Lanczos method through repeated matrix-vector multiplication \cite{calvetti1994implicitly}. Therefore, the time complexity of CLE and CLE* is $O(N^2)$. CPU times of the 13 algorithms are compared in Fig.~\ref{fig.3}. CRA and $\ddot{\mathbf{S}}^{CRA}$ take the longest times for large-scale networks, and the CPU time of $\ddot{\mathbf{S}}^{CRA}$ is slight longer than CRA (the difference is too small to differentiate in Fig.~\ref{fig.3}). Except CRA and $\ddot{\mathbf{S}}^{CRA}$, LO and the Katz index are the most time-consuming algorithms. Overall speaking, CLE and CLE* are competitive with the best-performed parameter-dependent algorithm LO, while require much shorter CPU time than global algorithms like LO and the Katz index.

\begin{figure}[ht!]
\setlength{\abovecaptionskip}{0pt}
\centering
	\includegraphics[width=1\textwidth]{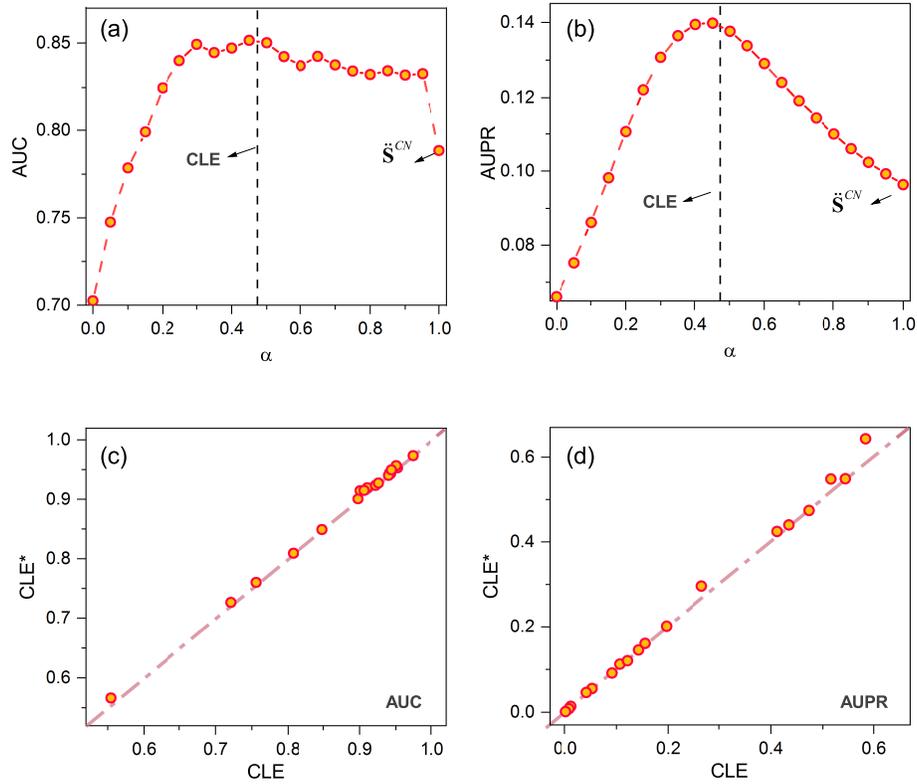}
    \caption{Comparison of CLE and CLE*. (a) and (b) show AUC and AUPR of CLE* versus $\alpha$ for DNC. (c) and (d) show the relation of CLE and CLE* on 18 real networks subject to AUC and AUPR. Each data point is averaged over 100 independent runs with $E^P$ containing 10\% links, and parameters of CLE* are tuned to their optimal values subject to the highest AUC values and the highest AUPR values respectively. As marked in the figures, CLE and $\ddot{\mathbf{S}}^{CN}$ correspond to $\alpha=0.474$ and $\alpha=1$, respectively. CLE and CLE* largely improve the prediction performance of $\ddot{\mathbf{S}}^{CN}$ by controlling the contribution of $\mathbf{v}_1$, and CLE performs almost the same to CLE* for all considered networks.}
\label{fig.4}
\end{figure}

We finally compare the parameter-free algorithm CLE and the parameter-dependent algorithm CLE*. To our surprise, CLE performs almost the same to CLE* on all considered networks. Taking DNC as a typical example, as shown in Fig.~\ref{fig.4}(a) and Fig.~\ref{fig.4}(b), AUC and AUPR of CLE are very close to CLE*, while AUC and AUPR of $\ddot{\mathbf{S}}^{CN}$ are much lower. Fig.~\ref{fig.4}(c) and Fig.~\ref{fig.4}(d) separately compare AUC and AUPR of CLE and CLE* on all 18 real networks. It is observed that almost all data points are located on diagonal lines with the mean difference between their AUC values being only 0.0025, and the mean difference between their AUPR values being only 0.0099. The Mann-Whiteny U Test \cite{mann1947test} also suggests that there is no significant difference between CLE and CLE* subject to AUC and AUPR ($p$-value $>0.05$).

\section{Conclusion}
This paper proposes two algorithms, CLE (parameter-free) and CLE* (param-eter-dependent), by controlling the leading eigenvector of the adjacency network $\mathbf{A}$. Experiments on 18 real networks show the following three main findings. (i) CLE outperforms the classical local similarity indices and their enhanced versions. (ii) CLE* is competitive with the global algorithm LO on accuracy, which is one of the most accurate algorithms thus far. (iii) The prediction performance of CLE and CLE* is highly competitive. In addition to the high prediction accuracy, CLE and CLE* have lower time complexity than many known global algorithms (e.g., LO and the Katz index) and thus can be applied at least to mid-sized networks.

\section*{Acknowledgments}
We acknowledge Dr. Carlo Vittorio Cannistraci for valuable discussion. This work was partially supported by the National Natural Science Foundation of China (Grant Nos. 11975071 and 61673086), the Science Strength Promotion Programmer of UESTC under Grant No. Y03111023901014006, and the Fundamental Research Funds for the Central Universities under Grant No. ZYGX2016J196.

\bibliography{mybibfile}
\end{document}